# Inverse designed broadband all-dielectric electromagnetic metadevices

**Authors:** Francois Callewaert[1], Vesselin Velev[2], Alan V. Sahakian[1], Prem Kumar[1,2] and Koray Aydin[1*]

[1]Department of Electrical Engineering and Computer Science, Northwestern University, Evanston, IL 60208

[2]Department of Physics and Astronomy, Northwestern University, Evanston, IL 60208

*Correspondence to: aydin@northwestern.edu

**Abstract:** We present an all-dielectric metadevice platform realized by combining an inverse electromagnetic design computational method with additive manufacturing. As opposed to conventional flat metasurface based devices that are composed of resonant building blocks resulting in narrow band operation, our design approach creates non-resonant, extremely broadband ($\Delta\lambda/\lambda > 25\%$) metadevices. We propose and demonstrate high-efficiency (transmission $> 60\%$), using thin ($\leq 2\lambda$) all-dielectric metadevices that are capable of polarization splitting, beam bending, and focusing. Experimental demonstrations are performed at millimeter-wave frequencies using 3D-printed metadevices, however our proposed platform can be readily applied to the design and fabrication of electromagnetic and photonic metadevices spanning microwave to optical frequencies.

Conventional optical elements that control the polarization, phase and amplitude of electromagnetic (EM) radiation such as lenses, polarizers, beamsplitters, and mirrors are typically engineered at a scale larger than the wavelength. Within the last two decades, significant amount of research has been devoted to understanding light-matter interactions, designing novel materials and electromagnetic devices with subwavelength features. Metamaterials, and more generally materials composed of nanostructures with subwavelength feature size, have emerged as a viable platform to manipulate electromagnetic radiation in unconventional manners (*1, 2*). In particular, photonic crystals(*3*) and negative-index materials(*4*) have been used to achieve sub-diffraction lensing(*5-7*). More recently, metasurfaces (*8-10*) have gained substantial interest due to their ability to perform optical functionalities such as lensing(*11*), holograms(*12, 13*) and beam shaping(*14*) within an extremely thin layer. Although the ability to control phase, amplitude and polarization using subwavelength-thick metasurfaces is a promising route towards building miniature optical devices, they suffer from several drawbacks prohibiting their potential in replacing conventional bulk optical elements. Initial metasurface designs utilized plasmonic metals that exhibit high optical losses and thus were of relatively low efficiency(*15*). Lossy metals can be replaced with high-index dielectric materials like amorphous silicon (*16*), but such metasurfaces often rely on Mie-type resonances that result in a narrow wavelength range of operation (*14, 17*).

Typical metasurface design starts with identification of an optical resonator with a well-defined geometrical shape, such as triangles(*18*), rectangles(*12, 19*), ellipses(*14*) or V-antennas(*8, 15*). Phase information is then calculated for various geometrical parameters such as radius, width, orientation, etc.(*16*). Metadevices based on ultrathin metasurfaces often yield a polarization dependent and narrow band optical response since their design relies on subwavelength optical resonators.

Here, we introduce an inverse electromagnetic method(*20-23*) to design high-efficiency (>60%), broadband (Δλ/λ > 25%), dielectric-based thin (≤ 2λ) electromagnetic metadevices overcoming the aforementioned limitations. Inverse design opens up the entire design space to enable metadevices with increased and enhanced functionalities. In order to demonstrate the feasibility of our inverse design approach, we use additive manufacturing and to print a low-loss polymer into a complex geometrical pattern. We demonstrate the design, fabrication and characterization of wavelength-scale metadevices for bending, polarization splitting and focusing of EM radiation at millimeter-wave frequencies.

We propose a free-space polarization splitter meta-grating (Fig. 1A) that bends parallel and perpendicular polarizations to opposite diffraction orders, as well as a meta-grating that bends both polarizations to the same diffraction order. We also propose a ~λ-thick flat metalens (Fig. 1B) that converts an incoming plane-wave into a cylindrical wave which focuses the radiation at a chosen focal point away from the metalens.

We use the objective-first algorithm (*22, 24, 25*) in which we treat the electromagnetic wave equation:

$$\min_{\varepsilon, E} \nabla \times \nabla \times E - w^2 \varepsilon E$$

as an optimization problem, and solve alternatively for the electric field, **E** and the dielectric permittivity, $\varepsilon$, using the desired input and output electromagnetic field distributions. Such an optimization problem is non-convex; therefore, there is no general method to find the optimum solution. However, suitable solutions that satisfy desired functionality with acceptable performance can be reached. An on-chip wavelength splitter (*22*) and an optical diode (*26*) have been successfully demonstrated using such inverse electromagnetic design approach.

Bending and polarization splitting are achieved using meta-gratings that convert an input plane wave to an output plane wave having a different diffraction order than $m = 0$, with periodic boundary conditions along the *x*-axis. For metalenses, we aim to focus an input plane wave at a desired focal distance; hence, the output is chosen to be a cylindrical wave centered at a specific location. Metalenses do not perform like a grating; therefore, we set the boundary conditions of a perfectly matched layer (PML) along the *x* direction. Our inverse design is two-dimensional; therefore, we assume that the metadevices have infinite height along *z*. In practice, the fabricated devices are $\approx 10\lambda$ thick.

Inverse-designed metadevices are fabricated using additive manufacturing, commonly called 3D-printing. This bottom-up approach allows the fabrication of very complex devices with a large aspect ratio. Furthermore, 3D-printing is an incredibly scalable method, with resolutions ranging from 100 nm to 1 mm (*27*), allowing the fabrication of electromagnetic devices for applications from the visible to the millimeter- wave and microwave regimes (*28-31*). Here, we show that the proposed devices can be practically realized for microwave and millimeter-wave operation with polymer-based 3D printing approach (details in supplementary information).

First, we demonstrate a free-space polarization splitter operating at millimeter-wave frequencies. The proposed metadevice (Fig. 2A) deflects a normal incident plane-wave polarized along *y* (parallel) and *z* (perpendicular) directions into *m*=+1 and *m*=-1 diffraction orders respectively with high efficiency and over a broad bandwidth. We simulate the device with a periodic boundary condition along the y-direction and assume infinite thickness in the z-direction. The width is chosen to be ~$2\lambda$ to reach desired phase change as explained in the supplementary information. The periodicity, $L$ along $y$ is determined by the deflection angle $\theta$ of the desired diffraction order $m$ (here $m = \pm 1$ for all devices), following the grating equation $L\sin\theta = m\lambda$. We designed and optimized the metadevice for an operation frequency of 33 GHz (free space wavelength of $\lambda$=9.1 mm) and a deflection angle of $\theta = \pm 30°$, for which $L = 1.8$ cm. The inverse-design algorithm generates a binary refractive-index distribution of dielectric and air that is then printed with dimensions of 2 cm x 7.2 cm x 8 cm. A photograph of the 3D-printed metadevice is shown next to the computer-generated pattern in Fig. 2A, which shows the high fidelity of the 3D-printing method.

We measured the far-field angular transmission through the fabricated metadevice to verify the predicted polarization splitting behavior. Figure 2B plots the simulated and measured power distributions at 33 GHz. We observe that a plane wave with parallel polarization is bent at an angle $\theta = +30°$, whereas the perpendicular polarization is deflected with an angle of $\theta = -30°$. The total power transmitted by the metadevice at 33 GHz is measured to be 76% for the parallel polarization and 54% for the perpendicular polarization, which are lower than the simulated values of 90%. The discrepancy, we believe, is due to structure imperfections in the fabricated devices. The rejection ratio, defined as the ratio between the peak intensity and the maximum intensity outside the main peak, is experimentally found to be 5.2 dB and 7.0 dB for the parallel and perpendicular polarizations, respectively, which are close to the simulated values of 6.6 dB and 9.3 dB, respectively.

We perform full-field electromagnetic simulations to calculate the electromagnetic properties of the metadevice. We plot the vertical electromagnetic fields, i.e. $H_y$ for parallel polarization (Fig. 2C) and $E_z$ for perpendicular polarization (Fig. 2D), at 33 GHz (continuous-wave excitation is provided as a movie in the supplementary information). The spatial electromagnetic field distribution provides a clear picture of how the EM waves propagate inside the metadevice. The device presents a dielectric filling-fraction gradient along the y-direction from a part mostly filled with dielectric ($\varepsilon$=2.3) where the phase is shifted by 6π to a part mostly void ($\varepsilon$ = 1.0) where the phase is shifted by 4π only, creating a 2π phase shift in the y-direction. The polarization splitting is a result of the different phase-change response of the device to different polarizations owing to its complex dielectric shape.

Although we choose 33 GHz to be the frequency to optimize for highest efficiency in our inverse-design algorithm, we observe broad operation bandwidth that spans our entire measurement range, which is enabled by the inverse-design method favoring non-resonant dielectric structures. Figure 2 plots simulated (E,F) and measured (G,H) power transmitted in the far-field as a function of the angle and the frequency for parallel (E,G) and perpendicular (F,H) polarizations. The simulations and measurements agree well, apart for minor differences that can be explained by the finite number of periods in the printed structures as well as an imperfect plane-wave input.

In order to demonstrate the versatility and flexibility of the inverse-design approach, we designed and fabricated two additional metadevices that bend the millimeter-waves. The first one is a polarization splitter with a different bending angle of 15° (Fig. 3A). Similar to the 30° splitter, this metadevice presents a gradient of dielectric filling fraction along the y-direction with a larger periodicity ($L$ = 3.5 cm) in order to favor a smaller bending angle. The simulated and measured angular far-field transmitted powers are plotted for both polarizations at 33 GHz in Figure 3B. The measured rejection ratios for the 15° splitter are 8.2 dB and 10.6 dB for parallel and perpendicular polarizations respectively. In the case of perpendicular polarization, a wave propagating along the *y*-direction is created in the device (see Fif. S1B in the supplementary information), changing its behavior compared to that for parallel polarization. The designs, simulated fields and broadband far-field data are shown in Figure S1.

In addition to polarization beam-splitter, we also designed and realized a polarization-independent millimeter-wave bending metadevice (Fig. 3C) which bends both polarizations to the same diffraction order. Simulations and experimental results of the far-field power at 33 GHz are shown in Figure 3D for this polarization-independent beam bending metadevice, showing very good agreement between the theory and experiment. The designs, simulated fields and broadband far-field data are shown in Figure S2. Although polarization-independent bending of EM radiation can be achieved with a triangular blazed grating, such gratings deflect significant amount of power to higher diffraction orders (comparison provided in the supplementary information Fig. S3). On average from 26 to 38 GHz, the inverse-designed device reduces the amount of power sent into undesired diffraction orders by a factor of 2.8 for parallel polarization and 2.0 for perpendicular polarization when compared to a blazed grating of similar thickness.

As a next step, we propose and design flat metalenses using the inverse-design algorithm. Such metalenses focus an incoming plane wave onto a focal point, as illustrated in Figure 1B. We designed and fabricated two different metalenses with focal lengths of $2\lambda$ and $15\lambda$, respectively. Both lenses are optimized and scaled for operation around 38 GHz ($\lambda = 7.9$ mm). The first metalens is 1.5-cm wide, 10-cm long, the second is 2.5-cm wide and 15-cm long and both are 10-cm tall. A picture of each device is shown with the computer-generated design in Figure 4. The electromagnetic behavior of both these devices is simulated with a perpendicularly-polarized incoming plane wave. The electric-field intensity profiles for the short-range and long-range metalenses are plotted in Fig. 4A and 4B, respectively. The electric-field amplitude is shown for both devices in the supplementary information (Fig. S3, along with a movie). We also performed a 2D scan of the transmitted power behind the metalenses using a millimeter-wave probe antenna positioned at $z = 5$cm. The measured spatial intensity distribution in the $x$-$y$ plane for the short-range and long-range lenses are provided in Figs. 4C and 4D, respectively. The simulated and measured spatial-intensity distributions closely match, with small discrepancies, we believe, due to the imperfect nature of the plane wave input, and minor differences between the ideal designs and the fabricated devices. As expected, the first device focuses EM radiation 1.5 cm (~2$\lambda$) away from the device whereas the second device's focal point is located 12 cm (~15$\lambda$) away. The full-width-at-half-maximum (FWHM) of the focused radiation for both devices are 0.5 cm and 1.1 cm, respectively, corresponding to practical numerical apertures (NAs) of 0.8 and 0.36 respectively, close to the theoretical values of 0.82 and 0.53, respectively. The proposed devices also show broadband focusing behavior from 28 GHz to 40 GHz. We provide the measured and simulated intensity profiles for operation at 30 GHz in Fig. S5.

In conclusion, we have presented a platform for the design and fabrication of novel millimeter-wave metadevices that combines an inverse electromagnetic design algorithm with additive manufacturing. The proposed design and fabrication methodology can be generalized to other regions of the electromagnetic spectrum, including photonic devices, wherein the desired response can be defined in terms of input and output electromagnetic field distributions. Although we design and demonstrate metadevices in the millimeter-wave region, due to the scalability of Maxwell's equations, similar devices can be designed to operate in the visible to the microwave frequency range provided that low-loss dielectric materials can be additively fabricated with subwavelength feature sizes. The presented platform addresses the need for rapid versatile design and prototyping of compact, low-cost, low-loss, and broadband components that can be easily integrated into complex electromagnetic systems.

**Acknowledgments:** K.A. acknowledges financial support from the Robert R. McCormick School of Engineering and Applied Science at Northwestern University.

**Author contributions:** F.C. and K.A. proposed the idea and designed the experiments; F.C. developed the inverse-design algorithm, performed the numerical simulations, and fabricated the metamaterial devices using the 3D printing technique. F.C., V.V., P.K, and A.V.S. performed the microwave measurements; F.C. and K.A. analyzed the data; and F.C., P.K., A.V.S., and K.A. wrote the manuscript.

**FIGURES**

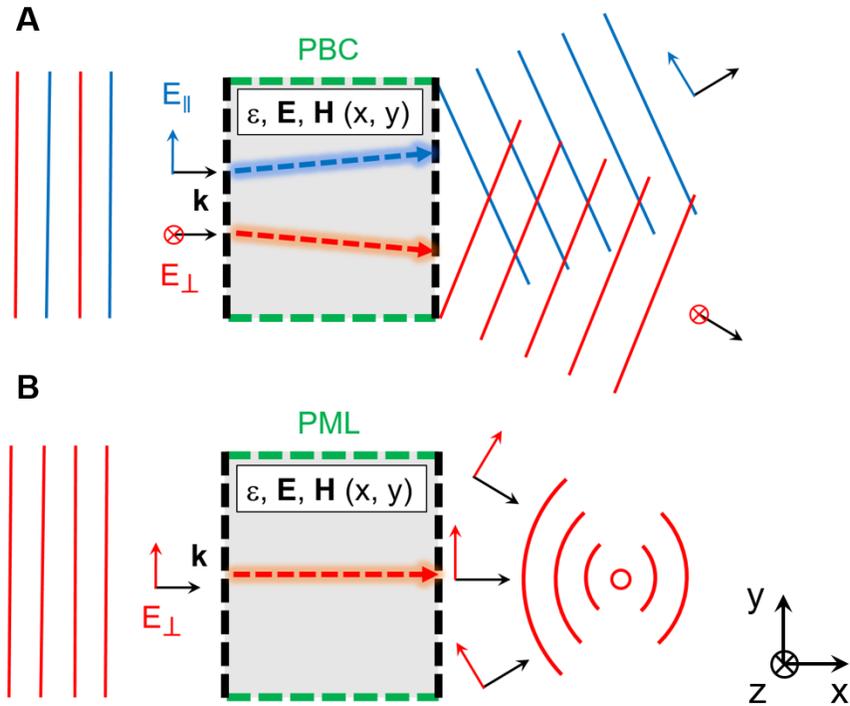

**Fig. 1.** Schematics for the inverse electromagnetic approach for designing free-space metadevices. The desired optical functionality is defined by a set of input and output conditions at the boundaries of the design space. A polarization splitter (A) is a grating that converts normally incident plane waves of parallel and perpendicular polarizations into two different diffraction orders. A flat metalens (B) is a device that converts a plane wave into a cylindrical wave converging to a chosen focal point.

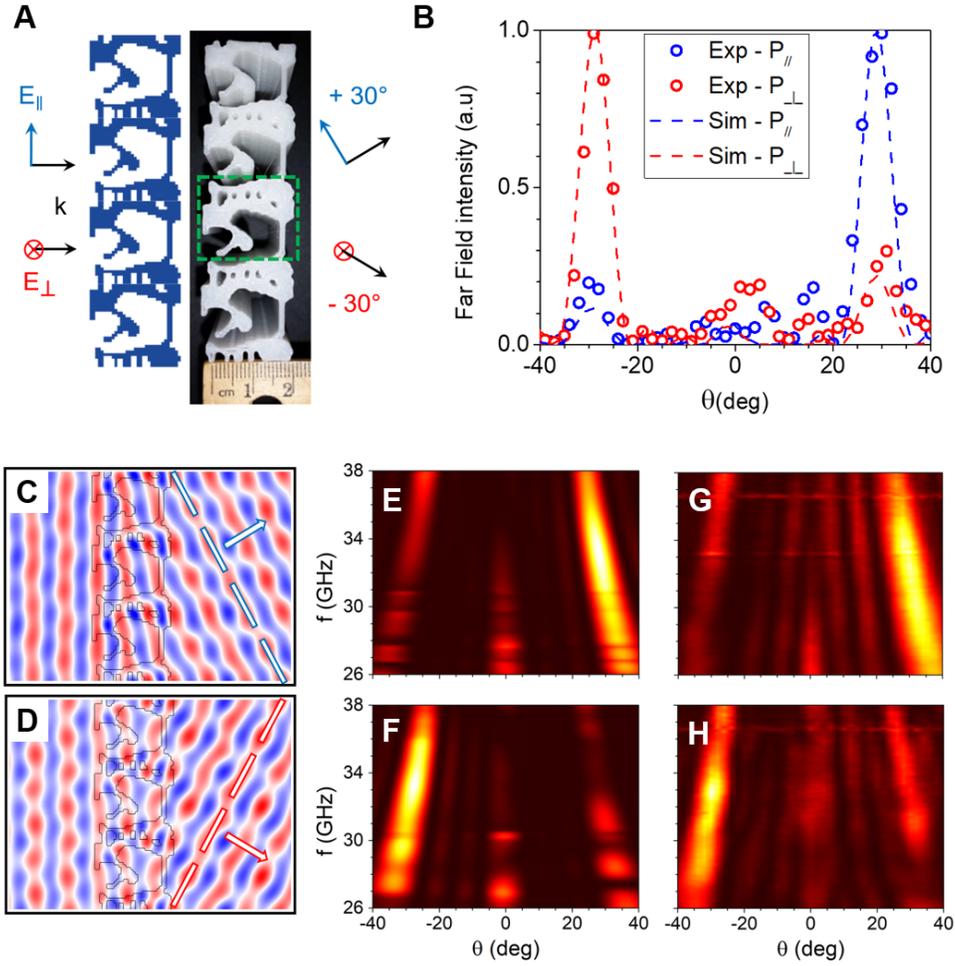

**Fig. 2.**

(A): Schematic drawing (left) and top-view photograph (right) of the 3D-printed 30° polarization splitter. The green rectangle indicates the unit cell of the grating. B) Simulated (dashed lines) and measured (circles) far-field power as a function of deflection angle for both parallel and perpendicular polarizations. (C) and (D): Simulated $\mathbf{H}_z$ and $\mathbf{E}_z$ field amplitudes for parallel (C) and perpendicular (D) polarizations, respectively, at 33 GHz. (E) to (H) Simulated (E, F) and measured (G, H) far-field intensity profiles as a function of the output angle and the millimeter-wave frequency for parallel (E, G) and perpendicular (F, H) polarizations.

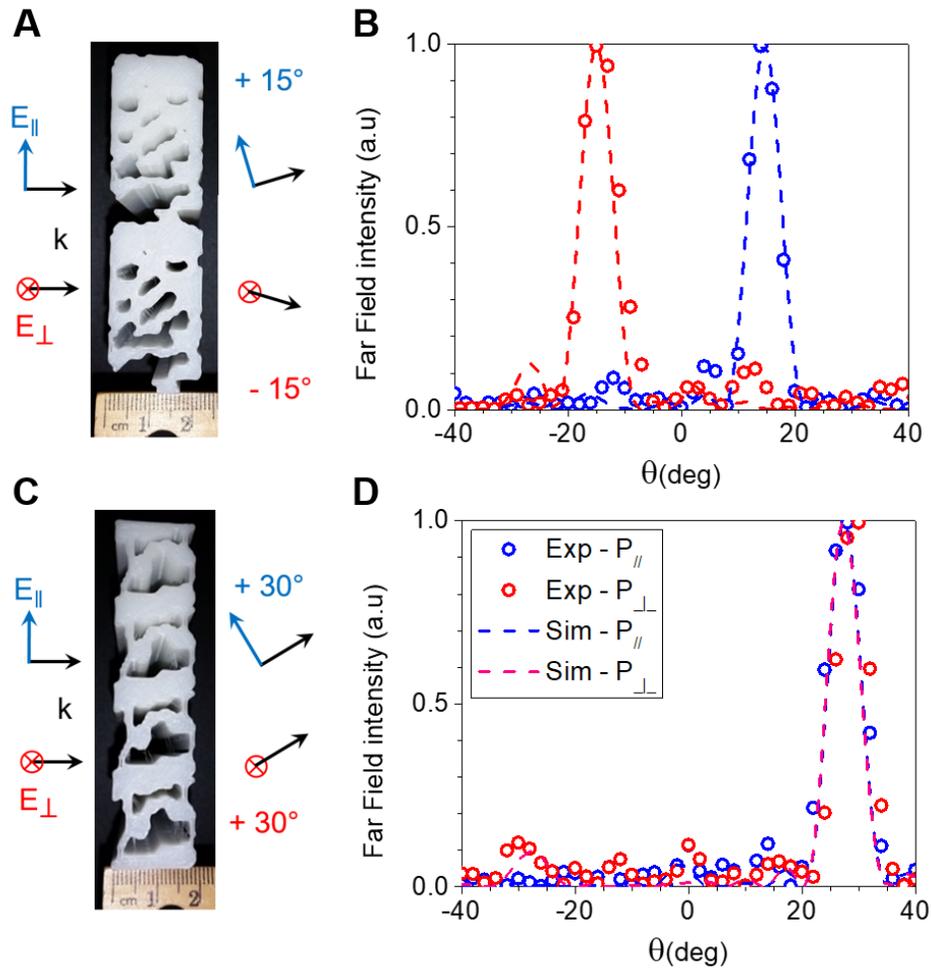

**Fig. 3.**

Photographs (A, and C) and simulated (dashed lines) and experimental (circles) far-field intensity plots of the 15° polarization splitter (B) and the 30° bend (D) as a function of the output angle for a frequency of 33 GHz.

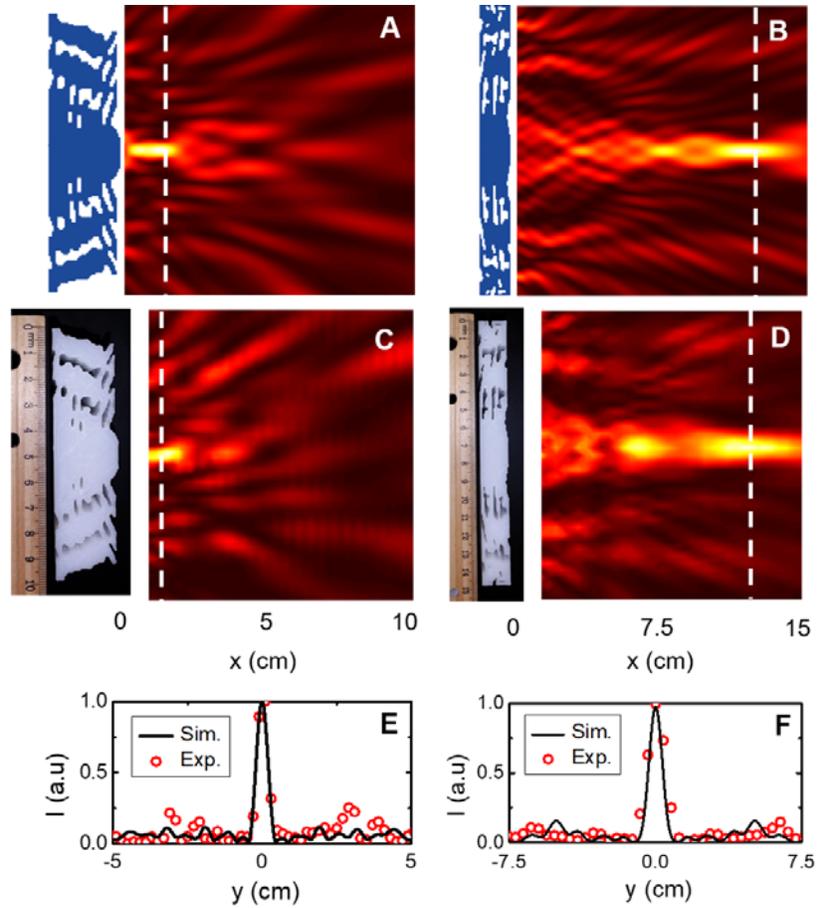

**Fig. 4.**

Simulated (A, B) and measured (C, D) spatial power distributions along the *x-y* plane from the output of the metalenses at 38 GHz. The input plane wave is generated by a horn antenna 1 m away on the left of the device while the output is measured with a probe antenna scanned along a 9x10 cm x-y plane for the first lens (A, C) and a 14 x 15 cm plane for the second lens (B, D). The first lens focuses perpendicularly polarized EM field 2λ away from the device whereas the second lens focuses it 15λ away. Schematics and pictures of the 3D-printed lenses are shown next to the simulated and experimental maps respectively. (E) and (F): Cross-section of the simulated (black line) and measured (red circles) power along the white dashed lines on the color maps for the first (E) and second (F) lens.

# Supplementary Materials for

## Inverse designed broadband all-dielectric electromagnetic metadevices

*Francois Callewaert[1], Vesselin Velev[2], Alan V. Sahakian[1], Prem Kumar[1,2] and Koray Aydin[1*]*

[1]Department of Electrical Engineering and Computer Science, Northwestern University, Evanston, IL 60208

[2]Department of Physics and Astronomy, Northwestern University, Evanston, IL 60208

**Materials and Methods**

The 3D-printed devices are made with high impact polystyrene (HIPS) and fabricated using a consumer grade 3D-printer based on fused deposition modeling. The material is chosen for its low cost and very low attenuation it presents to radiation in the microwave to millimeter-wave regtion, with a loss-tangent measured to be $\tan \delta < 0.003$ over the 26-38 GHz band. In this band, the real part of the dielectric constant of HIPS $\varepsilon' \approx 2.3$ ($n \approx 1.52$), which is then used as a constraint in our algorithm to inverse-design binary devices made of air ($\varepsilon = 1$) and HIPS ($\varepsilon = 2.3$). Because of the low index, the phase between the input and output is approximately proportional to the effective thickness of the polymer. Therefore, in order to allow a $2\pi$ phase shift between a part full of polymer and a part full of air, the device thickness needs to obey:

$$\Delta \phi = 2\pi (n-1) \frac{t}{\lambda} = 2\pi \times 0.52 \times \frac{t}{\lambda} \geq 2\pi,$$

which means that the thickness of the devices needs to be slightly larger than $2\lambda$.

To test the electromagnetic properties of the devices, a vector network analyzer (VNA) generates the input signal that is transmitted through a high-gain horn antenna placed far away from the sample (distance > 100λ) in order to produce a plane wave perpendicularly incident on the input surface. The device is surrounded by radar absorbing material to prevent the signal from going around it. For the first three devices, the transmitted power is measured in the far-field (>100λ) with a low-gain horn antenna as a function of the angle between –40° and 40° in 2° steps and as a function of the frequency between 26 GHz (11.5 mm) and 38 GHz (7.9 mm). For the lenses, the transmitted power is measured in the near-field with a probe antenna scanned in the x-y plane from the output of the devices through the focal region. The area scanned starts 1 cm away from the device due to technical constraints, which is the reason why the experimental intensity maps start at a hgher x-value than the simulated maps.

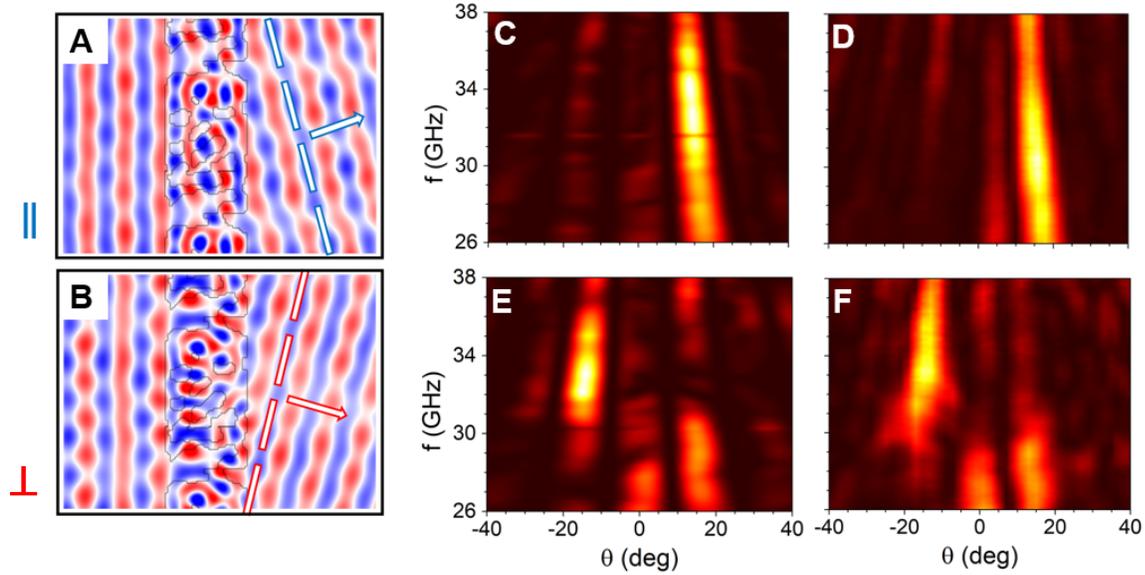

**Fig. S1.**

Simulated $\mathbf{H}_z$ (A) and $\mathbf{E}_z$ (B) fields in the 15° polarization splitter with a perpendicularly incoming plane wave for parallel (A) and perpendicular (B) polarizations and at a frequency of 33GHz. Simulated (C, E) and experimental (D, F) far-field intensity color maps as a function of the output angle between -40° and 40° and as a function of the frequency between 26GHz and 38GHz for both parallel (C, D) and perpendicular (E, F) polarizations.

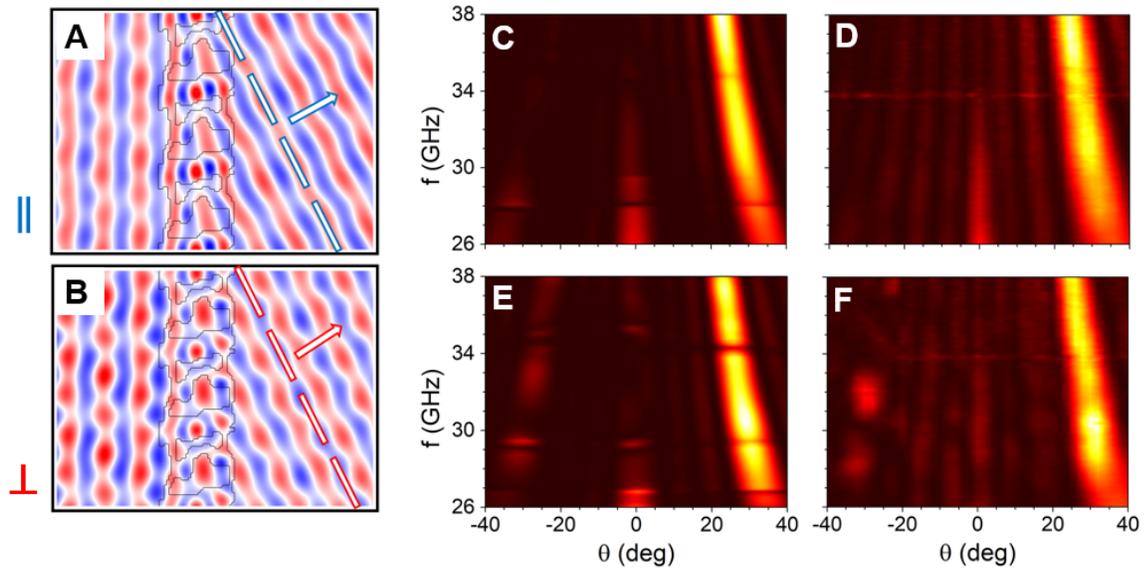

**Fig. S2**

Simulated $H_z$ (A) and $E_z$ (B) fields in the 30° bend with a perpendicularly incoming plane wave for parallel (A) and perpendicular (B) polarizations and at a frequency of 33GHz. Simulated (C, E) and experimental (D, F) far-field intensity color maps as a function of the output angle between -40° and 40° and as a function of the frequency between 26GHz and 38GHz for both parallel (C, D) and perpendicular (E, F) polarizations.

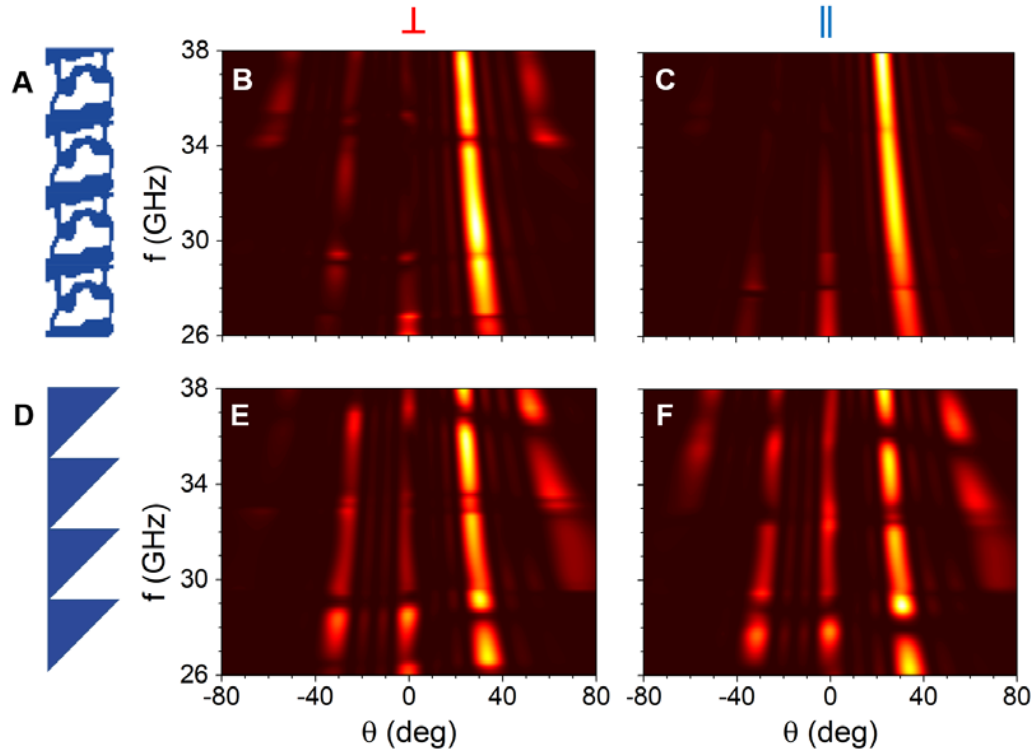

**Fig. S3**

Comparison between the performance of the inverse-designed device (A to C) and a blazed grating (D to F) optimized to bend electromagnetic radiation by 30° independently of the polarization. The simulated far-field intensities are represented for angles from -80° to 80° and for frequencies from 26GHz to 38GHz for perpendicular (B, E) and parallel (C, F) polarizations. As can be seen, the inverse-designed metadevice transmits a much lower power to undesired grating orders (23% for perpendicular polarization and 18% for parallel polarization) than the blazed grating (47% for perpendicular polarization and 51% for parallel polarization). Simulated rejection ratios at 32GHz are 10.1 dB and 12.4 dB for the inverse-designed bend, compared to 6.6 dB and 3.8 dB for the triangular grating for perpendicular and parallel polarizations respectively.

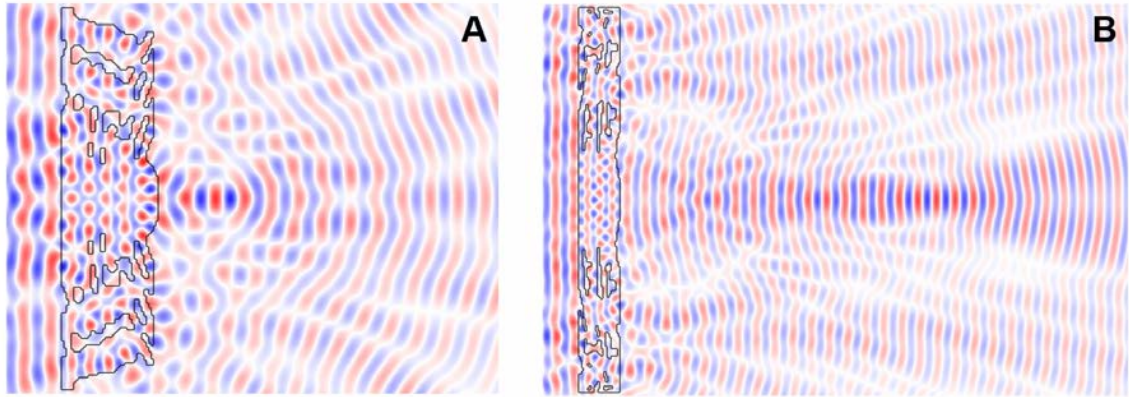

**Fig. S4**
Simulated $\mathbf{E}_z$ field amplitude color maps along the x-y plane at a frequency of 38 GHz, with the black lines showing the contour of the devices. (A) First lens with a focal distance of $2\lambda$; (B) second lens with a focal distance of $15\lambda$.

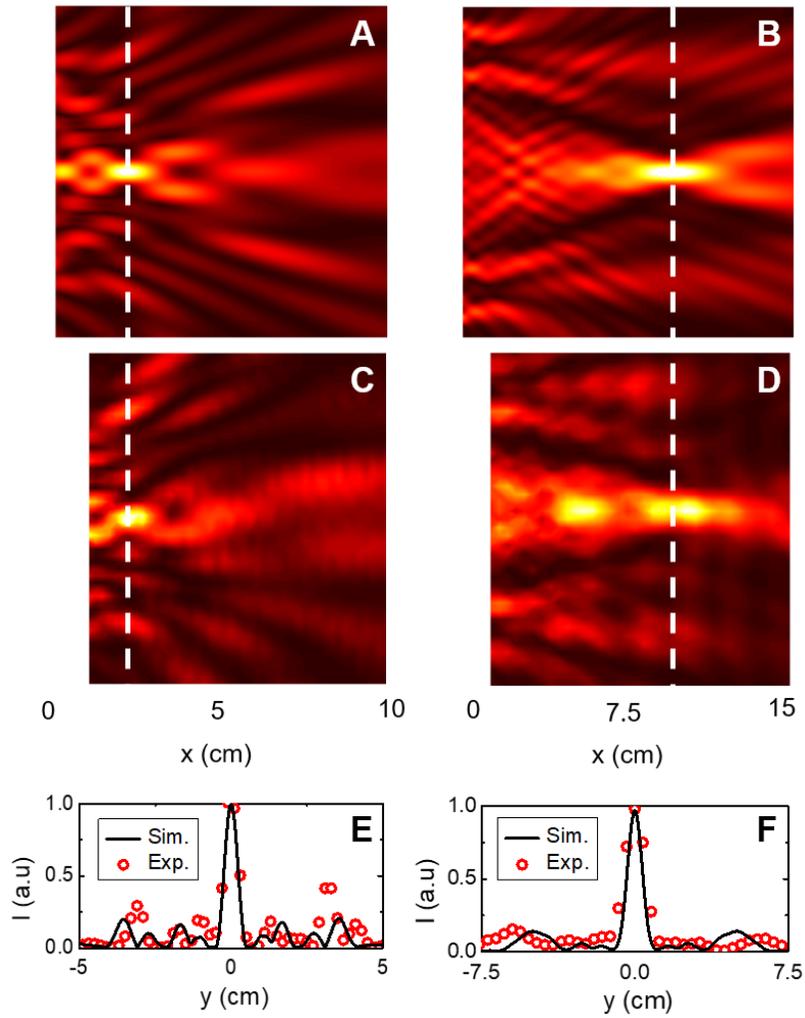

**Fig. S5**

Simulated (A, B) and experimental (C, D) electromagnetic intensity color maps along the x-y plane at the output of the devices at a frequency of 30 GHz for the first (A, C) and second (B, D) lenses. (E) and (F): Cross-section of the intensity along the white dashed lines on the color maps for the first